\documentclass[aps,prl,twocolumn,groupedaddress]{revtex4}
\usepackage{graphicx}
\input epsf
\usepackage{amsmath,amssymb}
\begin{document}
%\title{Blocking of interface traps in few layer MoS$_2$-on-SiO$_2$ FETs}
%\title{Interface traps' blocking transition and reversible control of threshold voltage in MoS$_2$ FETs}
\title{Blocking transition of interface traps in MoS$_2$-on-SiO$_2$ FETs}
%\title{Reversible control of threshold voltage in MoS$_2$ FETs by gate-field-cooling}
%\title{Reversible control of low temperature threshold voltage in MoS$_2$ FETs by cooling under gate-voltage}
\author{Santu Prasad Jana}
\affiliation{Department of Physics, Indian Institute of Technology Kanpur, Kanpur 208016, India}
\author{Suraina Gupta}
\affiliation{Department of Physics, Indian Institute of Technology Kanpur, Kanpur 208016, India}
\author{Anjan K. Gupta}
\affiliation{Department of Physics, Indian Institute of Technology Kanpur, Kanpur 208016, India}
\date{\today}

\begin{abstract}
Electrical conductivity with gate-sweep in a few layer MoS$_2$-on-SiO$_2$ field-effect-transistor shows an abrupt reduction in hysteresis when cooled. The hysteresis and time dependent conductivity of the MoS$_2$ channel are modeled using the dynamics of interface traps' occupancy. The reduction in hysteresis is found to be steepest at a blocking temperature near 225 K. This is attributed to the interplay between thermal and barrier energies and fitted using a distribution of the latter. Further, the charge stored in the blocked traps is programmed at low temperatures by cooling under suitable gate voltage. Thus the threshold gate-voltage in nearly non-hysteretic devices at 80 K temperature is reversibly controlled over a wide range.
\end{abstract}
\maketitle

\section{I: Introduction}
Single and few layer transition metal chalcogenides \cite{direct gap,TMDs} offer much potential for device applications including transistors \cite{direct gap1,how good} with high frequency capability \cite{high frequency}, logic gates \cite{logic,logic1} for integrated circuits \cite{ic} and optoelectronic \cite{photodetectors,optical helicity,light,photo1} devices. The MoS$_2$ single layer devices with direct band gap \cite{direct gap,direct gap1} in optical range have been of particular interest. The field effect transistors (FETs) based on MoS$_2$ show a very promising behavior with scalability, non-ideal behavior and degradation with time as the main hurdles. A pertinent culprit with interesting physics that contributes to non-ideal behavior is interface traps. Such traps lead to reduced mobility and response time as well as increased noise and hysteresis in transfer characteristics. Thus a more comprehensive understanding of the traps is necessary.

The phenomenon of blocking is common in magnetic systems. In ferromagnetic nano-particles, exhibiting superparamagnetism, \cite{superpara-blocking} the blocking arises from the interplay between an anisotropy energy barrier, the thermal energy and the Zeeman energy. As a result, the thermally activated switching rate ($\tau_{\rm s}(T)^{-1}$) between two magnetic states is a sharply rising function of temperature. Thus, their response to an applied magnetic field, measured over certain time ($\tau_{\rm m}$), shows hysteresis at low temperatures and non-hysteretic paramagnetic behavior at high temperatures. This crossover transition occurs at a blocking temperature $T_{\rm B}$ at which $\tau_{\rm s}(T_B)\approx\tau_{\rm m}$. In contrast, the blocking of traps in MoS$_2$ FETs leads to hysteresis reduction with cooling. A similar behavior is also observed in graphene FETs \cite{graphene-hyst}; though, the much sharper transfer characteristics in MoS$_2$ devices with a threshold gate-voltage help carry out a more quantitative analysis.

The threshold gate-voltage at which an FET shows a steep rise in conductance is controlled by both the traps' charge and the capacitive displacement charge across gate dielectric. A positive hysteresis in the transfer characteristics of MoS$_2$ FETs has been studied as a function of various parameters \cite{Bi-exponential,scalling behevior,interface,oxide traps} and attributed mainly to charge-traps. This arises from the traps that have a timescale comparable to the gate-voltage sweep-time. This also amounts to a relaxation in channel's conductance at varying time-scales. The fast traps do not lead to hysteresis but they shield the gate electric field restricting the density of mobile carriers in the channel. This broadens the threshold region and forbids the access to the ambipolar behavior in MoS$_2$ FETs even for gate voltages far exceeding the voltage equivalent to the energy gap. In addition, the electrostatic potential of the trap ions leads to reduced mobility of the channel carriers while the variation in the charge-state of traps gives rise to carrier density and mobility fluctuations.

In this paper, the transfer characteristics and its' time dependence in few layer MoS$_2$-on-SiO$_2$ FETs as a function of temperature is presented together with a model on the effect of traps on gate-dependent channel conductance. A trap's charge state determines the channel's chemical potential which in turn dictates the traps' occupancy. This makes it a complex non-linear system with coupling between traps' occupancy mediated by the channel. Thus, even the traps at a single energy and with the same barrier lead to non-exponential relaxation. The hysteresis and its temperature dependence is modeled using some simplifications and analogy with superparamagnets. Finally, the traps' blocking is used to reversibly control the threshold voltage at 80 K temperature.
\section{II: Experimental Details}
\begin{figure}[h]
	\centering
	\includegraphics[width=3.4in]{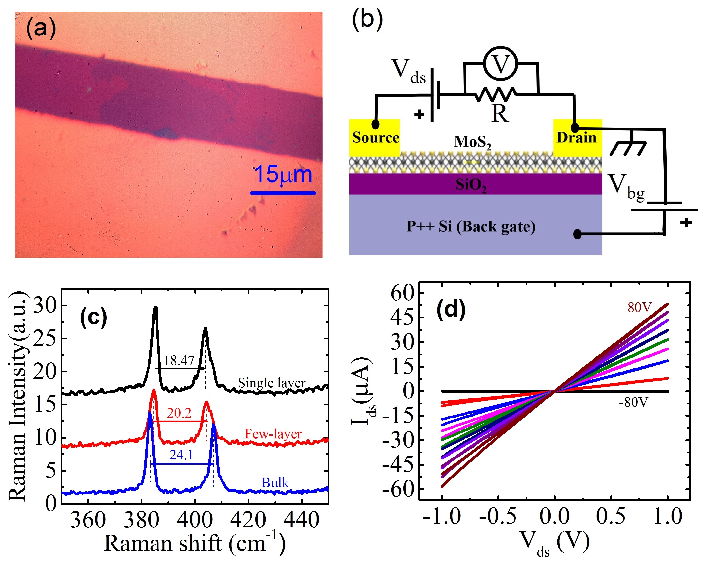}
	\caption{(a) Optical image of few-layer Mos$_2$ with gold contacts. (b) The electrical schematic drawing of MoS$_2$ FET. (c) Raman spectra measured on exfoliated single layer, few-layer and bulk MoS$_2$. (d) $I_{\rm ds}$ Vs $V_{\rm ds}$ for a few-layer device at different gate voltage values.}
	\label{fig:mos26}
\end{figure}

Few layer MoS$_2$ was transferred on SiO$_2$ by a dry method \cite{XYZ} from a natural MoS$_2$ single crystal (from SPI, USA) using commercial PDMS film based viscoelastic stamp. The latter is first fixed on a glass slide and an MoS$_2$ flake is transferred on it using a scotch tape. The mechanism of this transfer process uses the viscoelastic response of the PDMS film, which behaves as an elastic solid for a short time scale. So pulling the PDMS film from the scotch tape is done at high speed leading to strong adhesion of MoS$_2$ on PDMS as viscoelastic solid makes a strong conformal contact with the flake \cite{elastomeric stamp}. The PDMS with MoS$_2$ flake is aligned with a SiO$_2$/Si substrate fixed by carbon tape on a XYZ micro-manipulator and under an optical microscope. The stamp is removed with sufficiently low speed so that the adhesion of the flake to stamp is week and the flake gets transferred to the SiO$_2$ surface easily. Raman spectra, see Fig. \ref{fig:mos26}(c), were used to confirm few-layer nature of MoS$_2$.

The number of MoS$_2$ layers is determined by optical microscope contrast and verified by Raman Spectroscopy with 532 nm wavelength laser excitation. As seen in Fig.\ref{fig:mos26}(c) the separation between the  E$^1_{\rm 2g} $ and A$_{\rm 1g}$ Raman peaks is 18.47, 20.20 and 24.1 cm$ ^{-1}$, which correspond to the single-layer, few-layer and bulk MoS$_2$, respectively \cite{anomalous lattice vibration,optical identification}.
\begin{figure}[h]
	\centering
	\label{fig:mos27}
\end{figure}

We make 50 nm thick gold film source-drain contacts using mechanical masking with a 15 $\mu$m diameter tungsten wire. Use of Au without Cr/Ti adhesion layer promotes Ohmic contacts due to a very small difference in contact potentials of Au and MoS$_2$ \cite{Electrical-cont,Scandium}.  Mechanical masking avoids use of organic lithography resist which can leave residue on MoS$_2$. The wire is carefully aligned under an optical microscope with few-layer MoS$_2$ on SiO$_2$ substrate. Fig. \ref{fig:mos26}(a) shows an optical micrograph of a MoS$_2$ device with source-drain contacts. Two probe conductance down to 80 K temperature was measured, with the configuration shown in Fig. \ref{fig:mos26}(b) in a homemade vacuum cryostat with a heater for temperature control. A 10 k$\Omega$ series resistance was connected with the gate voltage supply, which was controlled by a data acquisition card using a LabView program. The Ohmic contacts were confirmed by two probe current-voltage characteristics as shown in Fig. \ref{fig:mos26}(d). The cryostat was pumped by a turbo-molecular pump to less than 10$ ^{-4} $ mbar pressure. When the cryostat is dipped into liquid nitrogen for cooling, the vacuum is expected to be much better than this. The device was annealed at 400 K in vacuum to minimize adsorbates on Mo$_2$ surface.

\section{III: Modeling of channel transport with time dependent interface traps}
In this section the temperature and time dependent channel transport in presence of interface traps is modeled. We first discuss the conduction in intrinsic 2D channel in presence of back-gate voltage. This is followed by details of the traps, their dynamics and their collective influence on the channel to model its time dependent transport and hysteresis. Finally we make some simplifications to model the temperature dependence of channel hysteresis and blocking transition of the traps.
\subsection{III-A: Intrinsic 2D semiconductor channel}
Consider an intrinsic 2D semiconductor with dispersion $E(k_{\rm x},k_{\rm y}) =\hbar^2(k_{\rm x}^2+k_{\rm y}^2)/2m^*$ on a gate-oxide with $m^*$ as the effective mass. This leads to an energy independent density of states (DOS) $g(E)=g_{\rm 2D}=g_{\rm s}g_{\rm v}m^*/2\pi\hbar^2$ with $g_{\rm s}$ as spin- and $g_{\rm v}$ as valley-degeneracy. The electron and hole densities, $n$ and $p$, respectively, will be given by $n=\int_{E_{\rm c}}^\infty g(E)f(T,E-\mu_{\rm ch})dE$ and $p=\int_{-\infty}^{E_{\rm v}} g(E)[1-f(T,E-\mu_{\rm ch})]dE$. Here $\mu_{\rm ch}$ is the chemical potential of the channel and $f(T,E)=[1+\exp{(E/k_{\rm B}T)}]^{-1}$ is the Fermi function. Eventually, the constant DOS in 2D leads to the expressions for $n$ and $p$ as
\begin{align}
n(T,\mu_{\rm ch})= &g_{\rm 2D}k_{\rm B}T\ln[1+\exp{\{\beta(\mu_{\rm ch}-E_{\rm c})\}}]\nonumber\\
p(T,\mu_{\rm ch})= &g_{\rm 2D}k_{\rm B}T\ln[1+\exp{\{\beta(E_{\rm v}-\mu_{\rm ch})\}}].
\label{eq:el-hole-den}
\end{align}

\begin{figure}[h]
	\centering
	\includegraphics[width=3in]{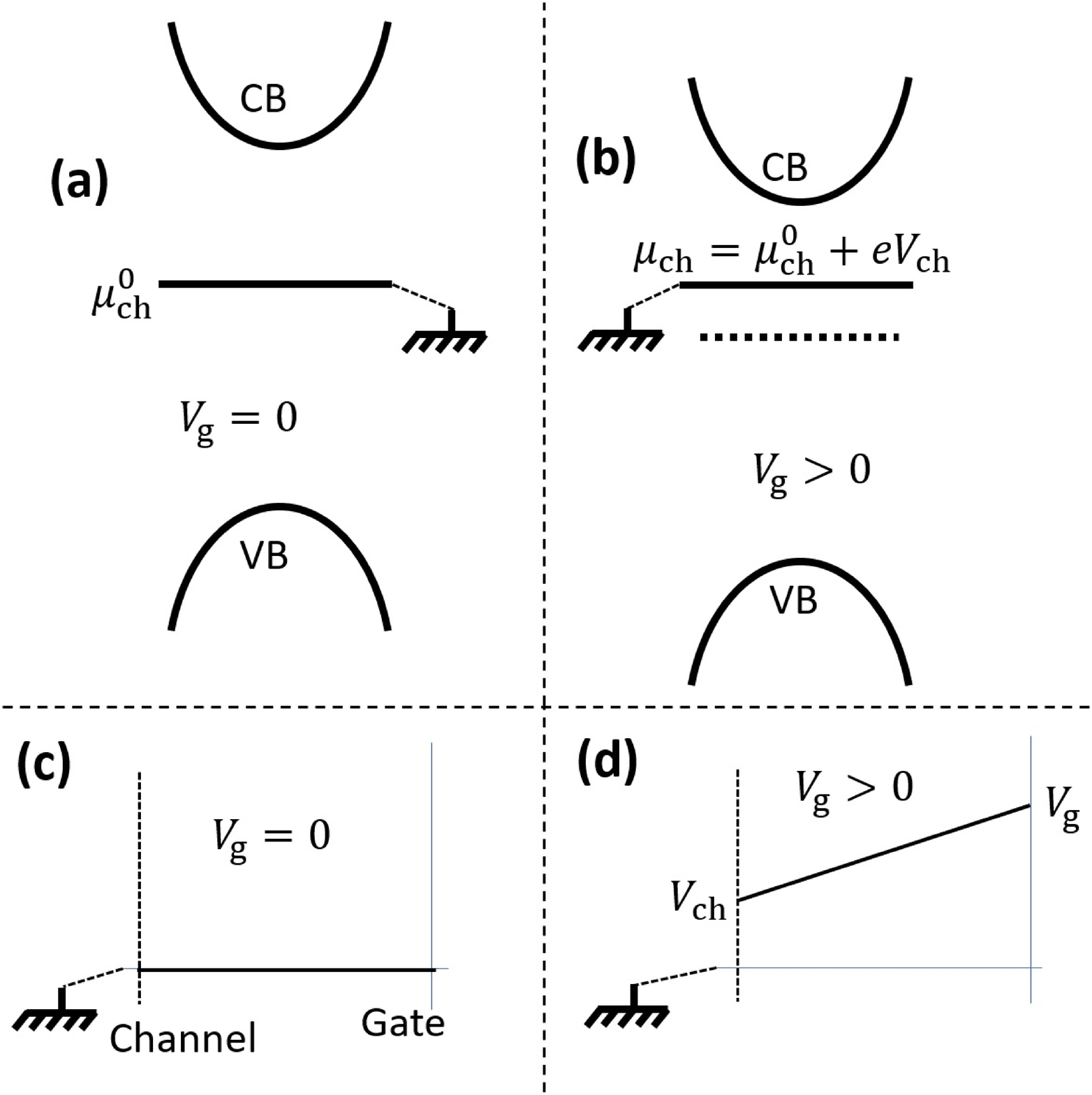}
	\caption{(a) and (b) show the channel bands while (c) and (d) show the electrostatic potential profile between the channel and the gate. (a), (c) are for $V_{\rm g}=0$ and (b), (d) for $V_{\rm g}>0$. The channel potential ($V_{\rm ch}$) rises with $V_{\rm g}$ and leads to a downward shift of the bands by $eV_{\rm ch}$ relative to the channel's chemical potential $\mu_{\rm ch}$ which is fixed to the ground reference. The jump by $V_{\rm ch}$ between drain (or source) and the channel, see (d), arises due to channel's quantum capacitance when the channel acquires a charge density.}
	\label{fig:Vg-band-shift}
\end{figure}

The net charge density in the channel $\sigma_{\rm ch}=e(p-n)$, in general, arises from dopants, gate electric field and interface trap charges. For an intrinsic 2D MoS$_2$ channel with zero gate electric field and no traps $\sigma_{\rm ch}=0$ and thus $n=p$ and so the chemical potential $\mu_{\rm ch}^0=(E_{\rm c}+E_{\rm v})/2$. This also assumes identical DOS for the valence and conduction bands. When a gate voltage $V_g$ is applied the channel potential changes to $V_{\rm ch}$ and $\mu_{\rm ch}=\mu_{\rm ch}^0+eV_{\rm ch}$, see Fig. \ref{fig:Vg-band-shift}. This leads to a non-zero $\sigma_{\rm ch}$ given by
\begin{align}
\sigma_{\rm ch} = -\gamma C_{\rm ox} \left(\frac{k_{\rm B}T}{e}\right) \ln\left[\frac{f(T,E_{\rm v}-\mu_{\rm ch}^0-eV_{\rm ch})}{f(T,\mu_{\rm ch}^0+eV_{\rm ch}-E_{\rm c})} \right].
\label{eq:sigma-Vs-Vch}
\end{align}
Here, the dimensionless $\gamma=e^2g_{\rm 2D}/C_{\rm ox}$ is the ratio of channel's quantum capacitance in the degenerate limit and the per-unit-area gate-oxide capacitance. The latter is given by $C_{\rm ox}=\kappa\epsilon_0/d$ with $d=300$ nm as SiO$_2$ thickness, $\kappa=4$ as its dielectric constant and $\epsilon_0$ as the permittivity of free space. The non-linear relation between $\sigma_{\rm ch}$ and $V_{\rm ch}$ in Eq. \ref{eq:sigma-Vs-Vch} amounts to a non-linear quantum capacitance \cite{Jena-2D-mater} of the channel. Note that a positive $V_{\rm ch}$ leads to a negative $\sigma_{\rm ch}$ and thus an increase in electron density. A positive $V_{\rm g}$ leads to a positive charge density at the gate electrode which will be equal and opposite to the combined charge densities of the channel carriers, dopant ions and trap ions.

As shown in Fig. \ref{fig:Vg-band-shift}(d), the overall applied $V_{\rm g}$, between drain/source, kept at ground potential, and the gate electrode, is shared between the channel potential $V_{\rm ch}$ and the voltage drop across the gate dielectric. For positive $V_{\rm g}$ and with $\mu_{\rm ch}$ as the fixed zero energy reference, the electron (negative charge) energy-bands of the channel shift downward by $eV_{\rm ch}$ leading to an increase in electron density in the channel. Thus $V_{\rm g}-V_{\rm ch}=-\sigma_{\rm ch}/C_{\rm ox}$, in the absence of dopants and traps. This and Eq. \ref{eq:sigma-Vs-Vch} lead to
\begin{align}
V_{\rm g}=V_{\rm ch}+\gamma \left(\frac{k_{\rm B}T}{e}\right) \ln\left[\frac{f(T,E_{\rm v}-\mu_{\rm ch}^0-eV_{\rm ch})}{f(T,\mu_{\rm ch}^0+eV_{\rm ch}-E_{\rm c})} \right].
\label{eq:Vg-Vs-Vch}
\end{align}
Assuming same mobility $\mu$ and DOS for e and h, the channel's conductivity $G=(n+p)e^2\mu$ is given by,
\begin{align}
G &=-e^2\mu g_{\rm 2D}k_{\rm B}T\times \nonumber\\ &\ln\left[f(T,E_{\rm v}-\mu_{\rm ch}^0-eV_{\rm ch})f(T,\mu_{\rm ch}^0+eV_{\rm ch}-E_{\rm c})\right].
\label{eq:cond-Vs-Vch}
\end{align}
\begin{figure}[h]
	\centering
	\includegraphics[width=3.4in]{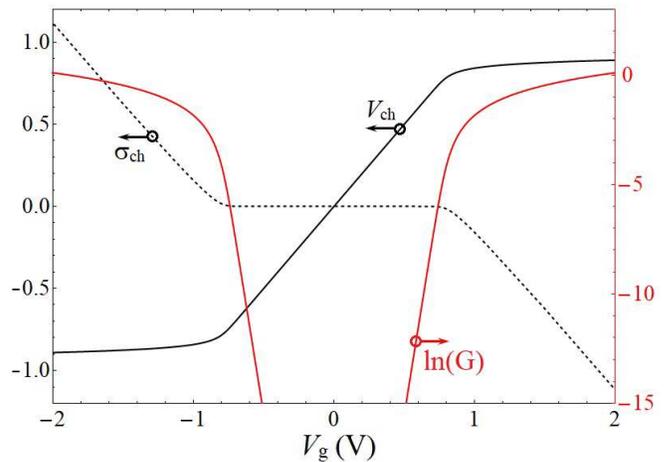}
	\caption{Variation with $V_g$ of $V_{ch}$ in Volts, channel charge density $\sigma$ in $\gamma C_{ox} (k_BT/e)$ units and $\ln{(G)}$ for no traps case. One can see that for this case the $eV_{th}$ values for channel conduction are close to the conduction and valence band energies.}
	\label{fig:no-trap-trans}
\end{figure}

For single-layer MoS$_2$ we use $g_{\rm s}=2$, $g_{\rm v}=1$ and $m^*=0.57m_{\rm e}$ to get $g_{\rm 2D}=2.64\times10^{14}$ eV$^{-1}$cm$^{-2}$ and with $d=300$ nm we get $\gamma=3570$. Further, for intrinsic MoS$_2$ channel we take $\mu_{\rm ch}^0=0$, $E_{\rm c}=1$ eV and $E_{\rm v}=-1$ eV. The $\sigma_{\rm ch}$, $V_{\rm ch}$ and $\ln(G)$, thus obtained, are plotted as a function of $V_{\rm g}$ in Fig. \ref{fig:no-trap-trans} using Eqs. \ref{eq:sigma-Vs-Vch}, \ref{eq:Vg-Vs-Vch} and \ref{eq:cond-Vs-Vch}, respectively, at $T=300$ K. We see that the channel conductivity and carrier density stay close to zero for $|eV_{\rm g}|<E_{\rm g}/2$ and rise abruptly beyond this range. Thus we expect the threshold voltages $eV_{\rm th}\sim \pm E_{\rm g}/2$. This is far from what is seen in actual experiments. We next discuss traps' energies, relative to channel chemical potential, and the barrier for charge exchange with the channel.

\subsection{III-B: Charge traps near 2D channel}
Traps have some similarities to dopants. Near room temperature a donor dopant exists in a semiconductor as a positively charged ion and a band-electron binds to it with an energy $E_{\rm d}$, just below $E_{\rm c}$. A negatively charged acceptor dopant, on the other hand, has a bound-hole state with an energy $E_{\rm a}$, just above $E_{\rm v}$. An ionized donor dopant leads to a mobile electron in the conduction band while keeping the overall system charge neutral. As dictated by the Fermi distribution, a bound state at $E_{\rm d}$ is unfavorable as compared to an electron at the chemical potential $\mu_{\rm ch}$ of the band-system as $\mu_{\rm ch}$ is close to the middle of the gap for low doping. Similarly, an acceptor dopant leads to a hole in the valence band keeping the overall system neutral and as compared to a hole at the chemical potential of the band system a hole-bound state at $E_{\rm a}$ is unfavorable.

The interface traps differ from dopants in three major ways. First, they are weakly coupled to the channel with an energy barrier between the trap- and channel-bound electron states. The barrier heights for different traps can be different. Small barrier heights will lead to a very small electron exchange time. Second, the energy for trap bound electron state need not be close to $E_{\rm c}$ and it can be anywhere relative to the bands. Third, a band-hole or band-electron will not be able to bind to a trap, and particularly so for large barriers or small coupling. The channel carriers will feel a weak electrostatic potential of the trap ion affecting its mobility. Further, a trap, assumed to be isolated in the sense of not interacting with other traps, will only form localized states and not extended band-like states.

\begin{figure}[h]
	\centering
	\includegraphics[width=3.4in]{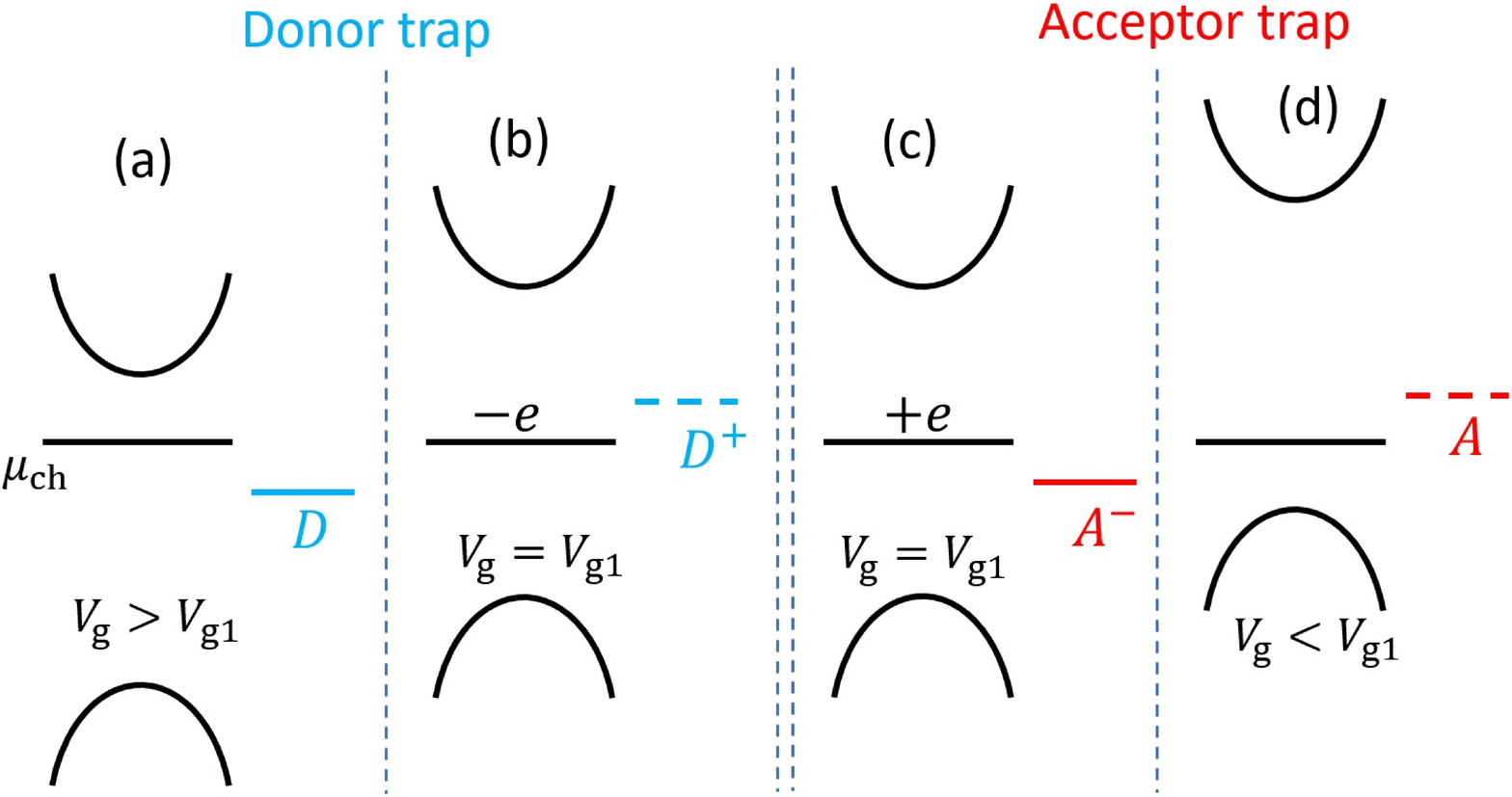}
	\caption{Schematics to illustrate the correlation between the equilibrium charge state of the donor and acceptor traps for different chemical potential values of the channel. The energies of the two traps relative to $\mu_{\rm ch}$ are dictated by the electrochemical reduction potentials as discussed in the text. When $V_{\rm g}$ is reduced [from (a) to (b) or (c) to (d)] leading to raising of bands (relative to $\mu_{\rm ch}$) and trap level, both the donor trap and acceptor trap, at the energies shown, lose an electron for the new equilibrium state.}
	\label{fig:traps}
\end{figure}
The relative energies of an electron bound to a trap versus that in the channel will be dictated by the electrochemical reduction potential of the two. A more positive reduction potential indicates an increased affinity for electron. For a donor trap $D$ if the reduction potential for the reaction $D^++e^-\rightarrow D$ exceeds the channel's reduction potential, \emph{i.e.} $Ch+e^-\rightarrow Ch^-$ then it is energetically favorable for donor trap to stay in neutral state. In the schematic band-energy diagram, see Fig. \ref{fig:traps}(a), this can be depicted as an electron-occupied neutral donor trap-level being lower than $\mu_{\rm ch}$ by its excess reduction-potential relative to the channel. This also means that if $\mu_{\rm ch}$ is decreased, see Fig. \ref{fig:traps}(b), it will become energetically favorable for the donor trap to transfer an electron to the channel and exist as $D^+$ \cite{trap-band-info}. The actual electron transfer can be slow depending on the barrier height between the channel and trap-bound states. Similar argument can be put forward for an acceptor trap. If the reduction potential for $A+e^-\rightarrow A^-$ exceeds that of $Ch^++e^-\rightarrow Ch$, then it is energetically favorable for the acceptor trap to be in $A^-$ state. Fig. \ref{fig:traps}(c),(d) depict the two energetically favorable scenarios for $\mu_{\rm ch}$ relative to trap energy. Over the limited accessible gate-voltage range the other higher ionization states of the traps may not be accessible.

In equilibrium, a donor trap with energy several $k_{\rm B}T$ lower than $\mu_{\rm ch}$ will remain un-ionized and will not contribute carriers to the channel. Thus, the traps that are far away in energy from the range of interest of $\mu_{\rm ch}$ will not change their charge state. A net charge density, due to such far-energy traps, will give rise to an equal and opposite charge density in the channel and it can be incorporated in the model through a fixed $\sigma_0$.

Further, the acceptor traps that can change their state in the relevant range of $\mu_{\rm ch}$ can be, for the modeling purpose, considered as donor traps by incorporating an appropriate change in $\sigma_0$. Suppose the areal density of acceptor and donor traps, respectively, at a given energy $E$ is $N_{\rm A}$ and $N_{\rm D}$. When a fraction $1-x$ and $x$, respectively, of these traps are ionized the total charge density in the traps will be $\sigma_0-e(1-x)N_{\rm A}+exN_{\rm D}$, \emph{i.e.} $\sigma_0-eN_{\rm A}+ex(N_{\rm A}+N_{\rm D})$. This can be interpreted as a fixed charge density $\sigma_0-eN_{\rm A}$ together with $N_{\rm A}+N_{\rm D}$ donor traps of which $x$ fraction is ionized. This will be indistinguishable, for modeling the channel conduction, from the actual state. Alternatively, we can simplify by turning the donors into acceptors with appropriate change in fixed charge; however, we adopt the former as a convention.

The interface traps, all donors, can thus be assumed to have certain energy dependent density of states. The other characteristic feature is the barrier for electron exchange between the trap and the channel. This determines the time scale over which the electron can transfer, between a trap state and the mobile channel states, by thermal activation or perhaps by tunneling.

\begin{figure}[h]
	\centering
	\includegraphics[width=3.4in]{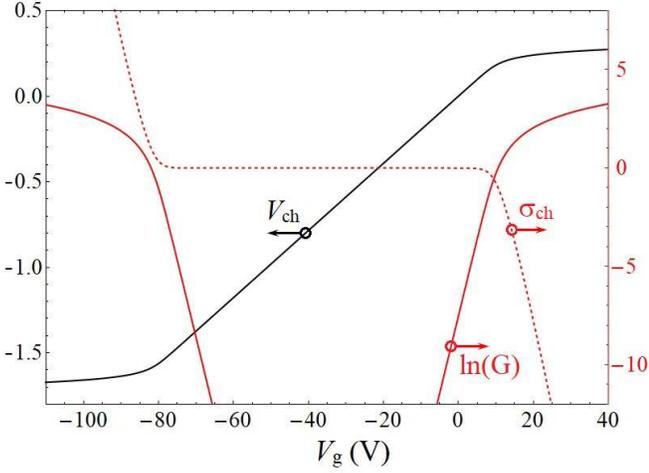}
	\caption{Variation with $V_g$ of $V_{ch}$ in Volts, channel charge density $\sigma$ in $\gamma C_{ox} (k_BT/e)$ units and $\ln{(G)}$ for $\gamma_{\rm ftr}=50$, corresponding to $g_{\rm ftr}=3.7\times10^{12}$ eV$^{-1}$-cm$^{-2}$, and $\mu_{\rm ch}^0=0.7$ eV. One can see that for this case the $eV_{th}$ values for channel conduction are much larger than the channel energy gap.}
	\label{fig:with-trap-trans}
\end{figure}
A simple model on the dynamics of trap-charge is presented in the next section while here we illustrate the equilibrium channel properties due to fast traps having a constant density of states $g_{\rm ftr}$. These traps act faster than the gate-voltage sweep time scale. Thus channel and traps exist in equilibrium for all $V_{\rm g}$ values where the (donor) traps with energy sufficiently below $\mu_{\rm ch}$ will be in neutral state while those above will be in the positively charged state. There is a fixed charge density $\sigma_0$ in the traps as discussed earlier. In such equilibrium when $V_{\rm g}$ is increased from zero, $\mu_{\rm ch}$ will change to $\mu_{\rm ch}^0+eV_{\rm ch}$ such that the interface traps in the energy range $eV_{\rm ch}$ will get neutralized from their positively ionized state, see Fig. \ref{fig:traps}(a) and (b). Therefore, the interface-traps' charge density will change from $\sigma_0$ to $\sigma_0-e^2V_{\rm ch}g_{\rm ftr}$. This will contribute an equal and opposite charge at the gate electrode. Thus Eq. \ref{eq:Vg-Vs-Vch} will change to
\begin{align}
V_{\rm g} = &(1+\gamma_{\rm ftr})V_{\rm ch}-\sigma_0/C_{\rm ox}\nonumber\\&+\gamma \left(\frac{k_{\rm B}T}{e}\right) \ln\left[\frac{f(T,E_{\rm v}-\mu_{\rm ch}^0-eV_{\rm ch})}{f(T,\mu_{\rm ch}^0+eV_{\rm ch}-E_{\rm c})} \right],
\label{eq:Vg-Vs-Vch-tr}
\end{align}
with $\gamma_{\rm ftr}=e^2g_{\rm ftr}/C_{\rm ox}$, \emph{i.e.} the ratio of traps' quantum capacitance to the gate capacitance. The other two equations, \emph{i.e.} Eq. \ref{eq:sigma-Vs-Vch} and \ref{eq:cond-Vs-Vch}, remain the same. Fig. \ref{fig:with-trap-trans} shows the calculated variation of $V_{\rm ch}$, $\sigma_{\rm ch}$ and $\ln(G)$ for an MoS$_2$ channel at $T=300$ K as a function of $V_{\rm g}$. This is for $\gamma_{\rm ftr}=50$, corresponding to $g_{\rm ftr}=3.7\times10^{12}$ eV$^{-1}$-cm$^{-2}$, and $\mu_{\rm ch}^0=0.7$ eV. Using Eq. \ref{eq:Vg-Vs-Vch-tr} with $V_{\rm ch}=0=V_{\rm g}$, the latter corresponds to static charge, $\sigma_0/e=g_{\rm 2D}k_{\rm B}T\ln[f(T,E_{\rm v}-\mu_{\rm ch}^0)/f(T,\mu_{\rm ch}^0-E_{\rm c})]=6.3\times10^7$ cm$^{-2}$.

A non-zero $g_{\rm ftr}$, thus, slows down the change in $V_{\rm ch}$ with $V_{\rm g}$ and also leads to an increase in the $V_{\rm th}$ values. This could be a reason for not seeing the hole doped transport regime over a large $V_{\rm g}$ range accessible in MoS$_2$ FETs. In the non-degenerate limit, \emph{i.e.} $(\mu_{\rm ch}-E_{\rm v}),(E_{\rm c}-\mu_{\rm ch})\gg k_{\rm B}T$, we use Eq. \ref{eq:cond-Vs-Vch} to get $d\ln(G)/dV_{\rm ch}=e/k_{\rm B}T$ for electron doping and Eq. \ref{eq:Vg-Vs-Vch-tr} to get $dV_{\rm g}/dV_{\rm ch}=1+\gamma_{\rm ftr}$. This leads to the expression for the subthreshold swing (SS) as an experimental method to find $\gamma_{\rm ftr}$, \emph{i.e}
\begin{align}
{\rm SS}=\left(\frac{d\log(G)}{dV_{\rm g}}\right)^{-1}=\frac{k_{\rm B}T\ln10}{e}(1+\gamma_{\rm ftr}).
\label{eq:sub-th-swing}
\end{align}

\subsection{III-C: Trap dynamics}
\begin{figure}[h]
	\centering
	\includegraphics[width=3.4in]{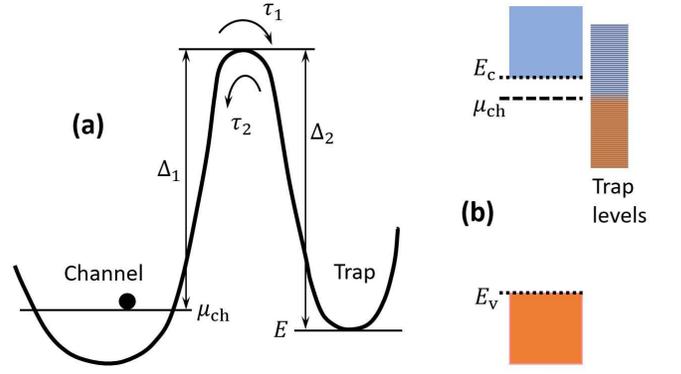}
	\caption{(a) Schematic of the barrier between the trap at energy $E$ and the channel filled with electrons up to its chemical potential $\mu_{\rm ch}$. (b) The trap levels relative to the bands. In equilibrium, the filled (orange) and empty (blue) states of the traps and the band are dictated by the Fermi function with chemical potential $\mu_{\rm ch}$.}
	\label{fig:trap-en-sch}
\end{figure}
The fast traps primarily increase the magnitude of $V_{\rm th}$ and SS. The slow traps, particularly the ones having a time scale comparable to $V_{\rm g}$ sweep time, are responsible for the positive hysteresis. The traps that are extremely slow to respond over $V_{\rm g}$ sweep time will only lead to a shift in $\sigma_0$. Fig. \ref{fig:trap-en-sch}(a) depicts a schematic where an electron can either be in the channel at $\mu_{\rm ch}$ or in the trap at energy $E$. There is a barrier between these two states whose height $\Delta_{1,2}$ will appear different from the two sides. This results into different transition rates $\tau_{1,2}^{-1}$ from the two sides. The time evolution of occupancy $p$, of an electron being in the trap at energy $E$, will be dictated by,
\begin{align}
\frac{dp}{dt}=-\tau_2^{-1}p+\tau_1^{-1}(1-p)=\tau_1^{-1}-(\tau_1^{-1}+\tau_2^{-1})p.
\label{eq:dpdt}
\end{align}
This leads to the solution
\begin{align}
p(t,E)=\frac{\tau_2}{\tau_1+\tau_2}+ \left[p(0,E)-\frac{\tau_2}{\tau_1+\tau_2}\right]e^{-(\tau_1^{-1}+\tau_2^{-1})t}.
\label{eq:p-vs-t}
\end{align}
Thus, at equilibrium, \emph{i.e.} $t\rightarrow\infty$, $p_{\rm eq}=\tau_2/(\tau_1+\tau_2)$.

Assuming identical attempt rates $\tau_{\rm a}$ from the two sides, we get $\tau_{1,2}=\tau_{\rm a} \exp{(\Delta_{1,2}/k_{\rm B}T)}$. With the barrier height difference $\Delta_2-\Delta_1=\mu_{\rm ch}-E$ we get $\tau_2/\tau_1=\exp{[(\mu_{\rm ch}-E)/k_{\rm B}T]}$. Thus $p_{\rm eq}$ can be written as,
\begin{align}
p_{\rm eq}(E)=\tau_2/(\tau_1+\tau_2)=[1+e^{(E-\mu_{\rm ch})/k_{\rm B}T}]^{-1},
\label{eq:trap-eqbm-occupancy}
\end{align}
which is the Fermi distribution $f(T,E-\mu_{\rm ch})$. Note that this is not an exact result and it will depend on the attempt rates from the two sides and on the degeneracy. In general a dopant-state occupancy is also not given by the exact F-D distribution \cite{Ashcroft-Mermin} due to the spin degeneracy of the dopant level.

The interface traps' areal charge density, dictated by traps' occupancy, will in turn determine the filling of the channel bands or $\mu_{\rm ch}$ value. We work in an independent electron approximation where the electron filling only affects $\mu_{\rm ch}$ and not the individual energies including channel-bands, traps, or the barrier $\Delta_2$. With change in $\mu_{\rm ch}$ the barrier $\Delta_1$ seen by the band electrons at $\mu_{\rm ch}$ will change but $\Delta_2$ will remain the same, see Fig. \ref{fig:trap-en-sch}(a). Thus eliminating $\tau_1$ in favor of $\tau_2$, $\mu_{\rm ch}$ and $E$, one can write Eq. \ref{eq:dpdt} and \ref{eq:p-vs-t}, respectively, as,
\begin{align}
\dot{p}=\frac{f(T,E-\mu_{\rm ch})-p}{\tau_2[1-f(T,E-\mu_{\rm ch})]},
\label{eq:pdot-2}
\end{align}
and
\begin{align}
p(t,E)=p&(0,E)e^{-t/\tau_2 f(T,\text{ } \mu_{\rm ch}-E)} \nonumber\\&+f(T, E-\mu_{\rm ch})[1-e^{-t/\tau_2 f(T,\text{ } \mu_{\rm ch}-E)}].
\label{eq:p-vs-t-1}
\end{align}
Here we assumed a time independent $\mu_{\rm ch}$ to solve for $p(t,E)$. However, $\mu_{\rm ch}$ will actually be determined by the time-dependent occupancy $p$ of various traps. We discuss this coupling between $p$ and $\mu_{\rm ch}$ next.
%Note that the probability of donor trap to be in positively ionized state will be $1-p$. We also note a few mathematical relations involving Fermi function as below:
%\begin{align*}
%f(T,-E)=1-f(T,E)\\
%\frac{d}{dE}\ln[f(T,E)]=-\frac{1-f(T,E)}{k_BT}.\\
%\end{align*}

\subsection{III-D: Time dependence of channel properties}
When the charge stored in interface traps changes with time, the density of mobile carriers, and thus $\mu_{\rm ch}$ will also change. The carriers in the channel are assumed to equilibrate over a time much smaller than $\tau_{1,2}$. We denote the areal density of states of slow traps with a transition time $\tau_2$ in range $\tau$ to $\tau+d\tau$ as $g_{str}(\tau,E)d\tau$. %For analyzing the time-dependent state of the channel, we also need to fix the $t=0$ state. This, in general, can be a non-equilibrium state depending on the history of $V_{\rm g}$; however, we assume it to be an equilibrium state and we choose $V_{\rm g}=0$ at $t=0$.

When $V_{\rm g}$ is changed from zero, the displacement charge and the fast traps will react immediately leading to a new $\mu_{\rm ch}$ and then the slow traps will start changing their occupancy. This will lead to a slow change in $\mu_{\rm ch}$. At an instant $t$ if the occupancy of the traps is $p(t,\tau,E)$ then the change in charge density of slow traps from the $t=0$ equilibrium state will be given by,
$\Delta\sigma_{\rm str}(t)=e\int\int[1-p(t,\tau,E)-1+p(0,\tau,E)]g_{\rm str}(\tau,E)d\tau dE$. By differentiating we get,
\begin{align}
\dot{\sigma}_{\rm str}=-e\int\int\dot{p}(t,\tau,E)g_{\rm str}(\tau,E)d\tau dE.
\label{eq:sig-sl-dot}
\end{align}
With slow traps, the $\sigma_0$ in Eq. \ref{eq:Vg-Vs-Vch-tr}, relating instantaneous $V_{\rm g}$ and $V_{\rm ch}$, gets replaced by $\sigma_0+\sigma_{\rm str}$. Differentiating this modified relation with respect to time, one gets,
\begin{align}
\dot{V_{\rm g}} &=(1+\gamma_{\rm ftr})\dot{V}_{\rm ch}-\dot{\sigma}_{\rm str}/C_{\rm ox}+\gamma \dot{V}_{\rm ch}\times\nonumber\\& \left[2-f(T,E_v-\mu_{\rm ch}^0-eV_{\rm ch})-f(T,\mu_{\rm ch}^0+eV_{\rm ch}-E_{\rm c}) \right].
\label{eq:Vg-Vs-Vch-tr-2}
\end{align}

A simple case is the trap states existing only at certain energy $E_0$ and with characteristic time $\tau_0$, \emph{i.e.} $g_{\rm str}(\tau,E)=n_{\rm str}\delta(\tau-\tau_0)\delta(E-E_0)$ with $n_{\rm str}$ as the areal density. This leads to $\dot{\sigma}_{\rm str}=-e\dot{p}n_{\rm str}$ with $\dot{p}$ dictated by Eq. \ref{eq:pdot-2} with $E=E_0-\mu_{\rm ch}^0$ and $\tau_2=\tau_0$. Using this in Eq. \ref{eq:Vg-Vs-Vch-tr-2}, we get
\begin{widetext}
\begin{align}
\dot{V}_{\rm ch}\left[1+\gamma_{\rm ftr}+\gamma\{1-f(T,E_v-\mu_{\rm ch}^0-eV_{\rm ch})+ f(T,E_{\rm c}-\mu_{\rm ch}^0-eV_{\rm ch})\}\right]=\dot{V_{\rm g}}-\left(\frac{en_{\rm str}}{\tau_0 C_{\rm ox}}\right)\frac{f(T,E_0-\mu_{\rm ch}^0-eV_{\rm ch})-p}{1-f(T,E_0-\mu_{\rm ch}^0-eV_{\rm ch})}.
\label{eq:Vg-Vs-Vch-tr-3}
\end{align}
\end{widetext}

Let's consider a step change in $V_{\rm g}$ from the $V_{\rm g}=0$ equilibrium state to $V_{\rm g1}$ at $t=0$ and then $\dot{V}_{\rm g}=0$ for $t>0$. $V_{\rm ch}$, $p$ and channel conductance $G$ will evolve with time with the latter being directly measurable. This evolution is dictated by two coupled first order ordinary non-linear differential equations, \emph{i.e.} Eqs. \ref{eq:pdot-2} and \ref{eq:Vg-Vs-Vch-tr-3}, that can be solved numerically.
\begin{figure}[h]
	\centering
	\includegraphics[width=3in]{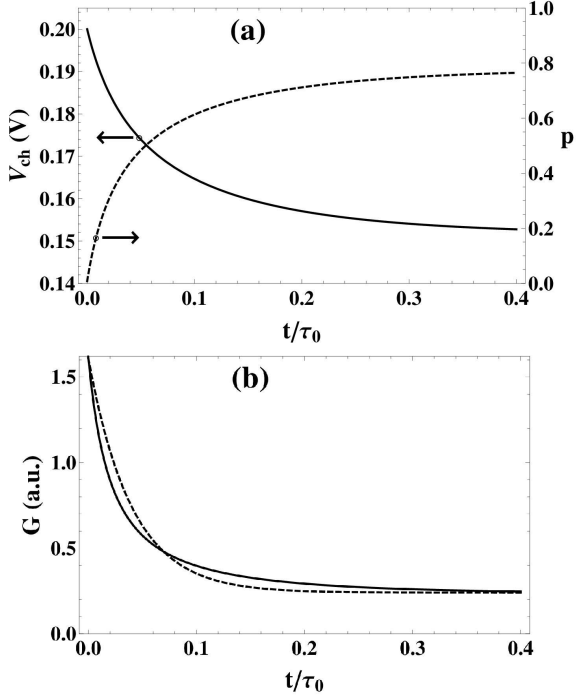}
	\caption{(a) shows the time evolution of channel voltage $V_{\rm ch}$ and trap occupancy $p$. The solid line in (b) shows the time evolution of conductance $G$ for a step change in $V_{\rm g}$ from zero to 21.4 V at $t=0$ (see text for details). The dashed line in (b) depicts an exponential decay function with characteristic time of 0.04$\tau_0$.}
	\label{fig:t-dep-ch}
\end{figure}
Figure \ref{fig:t-dep-ch} shows the time evolution at $T=300$ K of $V_{\rm ch}$, $p$ and $G$ for traps at single energy $E_0=0.82$ eV (from the middle of the gap) and when $V_{\rm g}$ is changed from zero to 21.4 V. We assume $\mu_{\rm ch}^0=0.7$ eV as arising from $\sigma_0/e=6.3\times10^7$ cm$^{-2}$. The other used parameters are: $n_{\rm str}=1.48\times10^{11}$ cm$^{-2}$ giving $(en_{\rm str}/C_{\rm ox})=2$ eV and $\gamma_{\rm ftr}=50$. The jump in $V_{\rm g}$ leads to a jump in $V_{\rm ch}$ from zero to 0.2 V and then it decreases continuously to about 0.15 V as the traps' charge increases. The channel Fermi energy $\mu_{\rm ch}=\mu_{\rm ch}^0+eV_{\rm ch}$ thus jumps to 0.9 eV and then decreases to about 0.85 eV. This is just above the trap energy of $E_0=0.82$ eV and thus leads to more than 75\% filling of the traps from nearly zero, see the discontinuous line in Fig. \ref{fig:t-dep-ch}(a). As seen in this figure none of the $V_{\rm ch}$, $p$ or $G$ time-evolution can actually be described by an exponential. This is illustrated in Fig. \ref{fig:t-dep-ch}(b) for $G$ where the dashed line shows an exponential relaxation with a characteristic rate $25\tau_0^{-1}$. This rate closely matches the initial relaxation rate, \emph{i.e.} $\tau_0^{-1} f(T,\mu_{\rm ch}-E_0)$ in Eq. \ref{eq:p-vs-t-1}, which, with $\mu_{\rm ch}^0-E_0=80$ meV, works out as $23.2\tau_0^{-1}$ at room temperature.

\subsection{III-E: Conductance hysteresis and blocking transition}
On SiO$_2$ the few layer MoS$_2$ is experimentally observed to be n-doped with its $\mu_{\rm ch}$ close to $E_{\rm c}$. Thus $V_{\rm th}$ for n-type conduction is usually found within $V_{\rm g}=\pm50$ V and p-type conduction is not observed. In the absence of slow traps one does not expect significant hysteresis in transfer characteristics. In such a case when $V_{\rm g}$ is ramped forward from an extreme negative to extreme positive value the n-type conduction will start at a threshold $V_{\rm g}$ value, say $V_{\rm th0}$ and it would stop at the same value when $V_{\rm g}$ is ramped back.

\begin{figure}[h!]
	\centering
	\includegraphics[width=2.7 in]{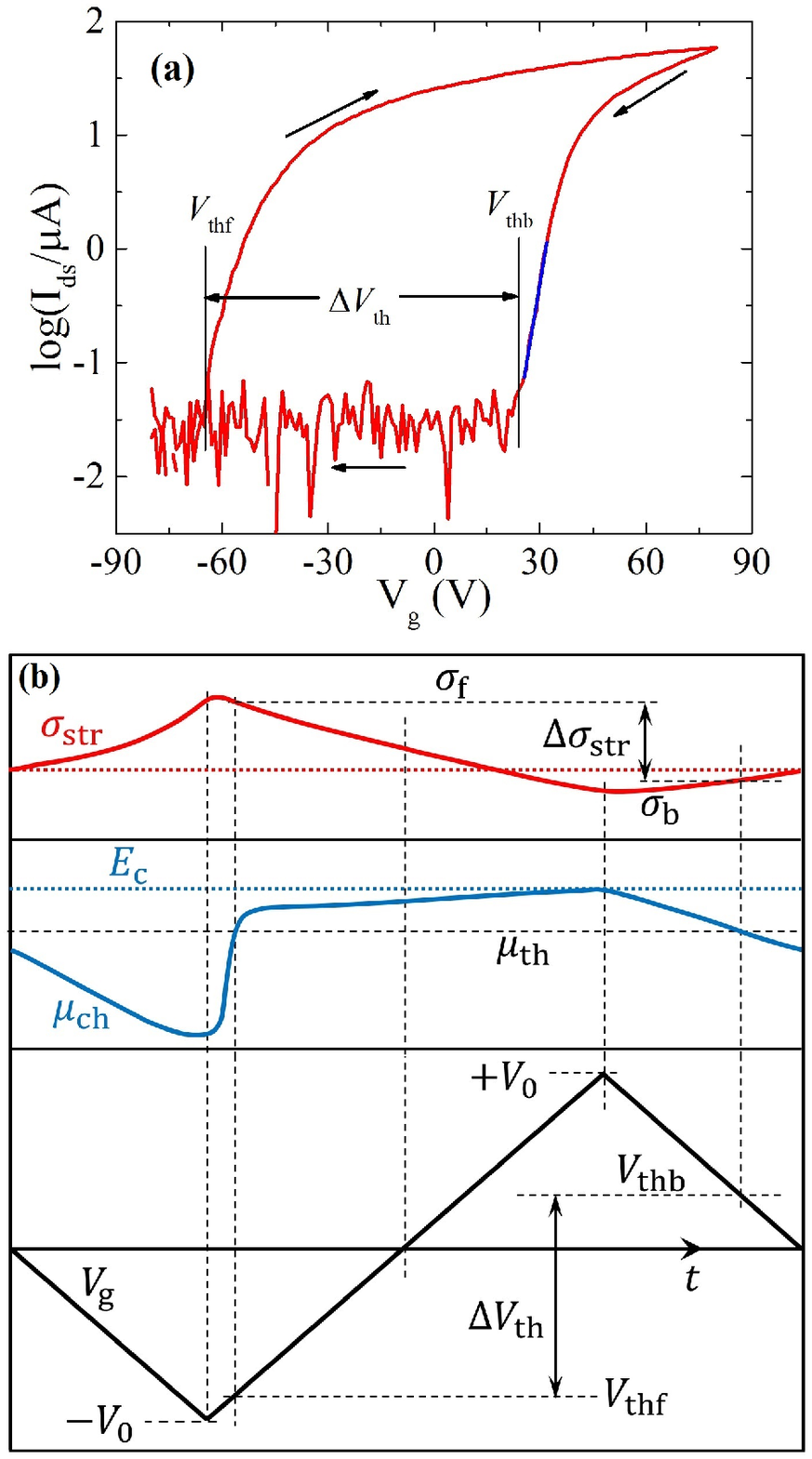}
	\caption{(a) shows the measured gate dependent drain current for a few layer MoS$_2$ at $V_{\rm ds}=1$ V as a function of $V_{\rm g}$ and over a $V_{\rm g}$ cycle from 0 to -80V, then to +80 and back to zero. (b) shows the schematic changes in slow-trap charge density $\sigma_{\rm str}$ (solid red line), $\mu_{\rm ch}$ (solid blue line) over a cyclic change in $V_{\rm g}$ (solid black line) between $\pm V_0$. The discontinuous red line depicts $\sigma_{\rm str}$ at initial and final zero $V_{\rm g}$ and the discontinuous blue line shows the $E_{\rm c}$ relative to which $\mu_{\rm ch}$ changes. The discontinuous horizontal black line just below $E_{\rm c}$ shows the $\mu_{\rm th}$ at which the channel starts conducting.}
	\label{fig:hyst-schem}
\end{figure}

Fig. \ref{fig:hyst-schem}(a) shows a measured conductance hysteresis loop at room temperature when $V_g$ is changed from 0 to -80 V then to +80 V and finally back to zero, all at the same rate. Fig. \ref{fig:hyst-schem}(b) is the schematic of how the slow traps charge density $\sigma_{\rm str}$ and $\mu_{\rm ch}$ change during this $V_{\rm g}$ cycle. When $V_{\rm g}$ is ramped to $-V_0$ from zero over certain time $\tau_{\rm m}$, both the channel and slow traps accumulate positive charge leading to a rise in $\sigma_{\rm str}$ and lowering of $\mu_{\rm ch}$ relative to $E_{\rm c}$. Now when $V_{\rm g}$ is ramped forward towards $+V_0$, the displacement charge in the channel changes fast while $\sigma_{\rm str}$ turns around slowly. This leads to a sharp rise in $\mu_{\rm ch}$ which crosses $\mu_{\rm th}$ at some threshold $V_{\rm g}=V_{\rm thf}<V_{\rm th0}$ when $\sigma_{\rm str}=\sigma_{\rm f}$. By this point only some traps change back their charge state to negative and the remaining still contribute mobile electrons in the channel together with those due to rising $V_{\rm g}$. At $V_{\rm g}=+V_0$ the traps accumulate a negative charge density and some of it, say $\sigma_{\rm b}$, will still remain when $V_{\rm g}$ turns around to reach $V_{\rm thb}>V_{\rm th0}$ at which the conduction stops.

At the conduction threshold, the channel carrier density, the chemical potential, and thus the quantum capacitance as well as the charge stored in fast traps will be same and independent of the $V_{\rm g}$ history. Therefore, the difference in charge density at the gate electrode for the two threshold $V_{\rm g}$ values will be equal to the difference in the charge density of the traps, \emph{i.e.} $\sigma_{\rm f}-\sigma_{\rm b}=C_{\rm ox}(V_{\rm thb}-V_{\rm thf})$ or $\Delta\sigma_{\rm str}=C_{\rm ox}\Delta V_{\rm th}$. Our objective here is to understand and model the temperature dependence of $\Delta V_{\rm th}$ which is proportional to $\Delta\sigma_{\rm str}$. The physics of this is similar to the super-paramagnet hysteresis, which is briefly discussed in the Appendix with the parameters that are relevant for traps. In high electron mobility transistors based on semiconductor hetro-junction the hysteresis, similar to superparamagnets, is found only at low temperatures \cite{hemt-hyst}, presumably due to a better coupling of the traps to the channel.

When one ramps $V_{\rm g}$ to $+V_0$, a given trap's occupancy will change according to Eq. \ref{eq:dpdt} with a time dependent $\mu_{\rm ch}=\mu_{\rm ch}^0+eV_{\rm ch}$ dictated by Eqs. \ref{eq:sig-sl-dot} and \ref{eq:Vg-Vs-Vch-tr-2}. This makes the occupancy of different traps coupled and rather complex. $\Delta V_{\rm th}$ is dictated by the difference in occupancy of the traps at the two threshold voltages. In the absence of the detailed knowledge about the traps' distribution, \emph{i.e.} $g_{\rm str}(\tau,E)$, and associated barriers we make certain simplifying assumptions. The magnitude of hysteresis and large SS value imply that the overall, fast and slow, trap density is large. We assume that this makes the overall change in $\mu_{\rm ch}$ much smaller than the energy gap $E_{\rm g}$ as well as the energy barrier $\Delta_2$. This also implies that only the traps in a narrow energy range, as compared to $E_{\rm g}$ and $\Delta_2$, change their state over experimental $V_{\rm g}$ sweep range.

For the temperature dependence $\Delta V_{\rm th}$, the exact details of $V_{\rm g}$ cycle will not make a significant difference as long as the overall time scale of the cycle is the same. We thus consider a case where $V_{\rm g}$ is first kept at zero for long enough time to achieve an equilibrium occupancy of traps and then it is abruptly changed to $-V_0$ and held at this value for time $\tau_{\rm m}$ and then it is ramped to zero over time $\tau_{\rm m}$. In this way $\mu_{\rm ch}$ will first decrease to certain lowest value $\mu_{\rm ch-}$ and then rise passing through $\mu_{\rm th}$ at certain $V_{\rm g}=V_{\rm thf}$ where the channel starts conducting. We consider a similar excursion from $V_{\rm g}=0$ equilibrium state, where the channel is insulating, to $+V_0$ where it is kept for $\tau_{\rm m}$. The channel will start conducting during this $\tau_{\rm m}$ when $\mu_{\rm ch}$ rises above $\mu_{\rm th}$ and reaches a maximum value $\mu_{\rm ch+}$. When $V_{\rm g}$ is ramped back to zero over time $\tau_{\rm m}$, $\mu_{\rm ch}$ will now go below $\mu_{\rm th}$ at certain $V_{\rm g}=V_{\rm thb}$ where the channel stops conducting.

We write from Eq. \ref{eq:p-vs-t-1} for the difference in occupancy of a trap at energy $E$, \emph{i.e.} $\Delta p=p_+-p_-$, corresponding to the two extreme gate voltages $\pm V_0$ as,
\begin{align}
\Delta p &=f(T,E-\mu_{\rm ch+})\left[1 -\exp{\left(-\frac{\tau_{\rm m}}{\tau_2f(T,\mu_{\rm ch+}-E)}\right)}\right]\nonumber\\ -&f(T,E-\mu_{\rm ch-})\left[1-\exp{\left(-\frac{\tau_{\rm m}}{\tau_2f(T,\mu_{\rm ch-}-E)}\right)}\right].
\end{align}
Here, we have assumed identical initial occupancy of the traps before $V_{\rm g}$ is brought to the two extreme values. Note also that $\tau_2=\tau_{\rm a}\exp(\Delta_2/k_{\rm B}T)$ is a steep function of temperature $T$ in the range of interest.

As discussed earlier, the traps that actually change their charge state in response to $V_{\rm g}$ will be in a narrow energy range. We believe that the potential barrier for an interface trap to change its charge state is actually much higher, \emph{i.e.} $\Delta_1,\Delta_2>>k_{\rm B}T,|\Delta_1-\Delta_2|$. So the temperature dependence of $\tau_2f(T,\mu_{\rm ch}-E)$ is primarily dictated by $\Delta_2$. In fact, the distribution in $\Delta_2$ or $\tau_2$ as compared to that in $E$ of relevant slow traps dominates the behavior. Thus we absorb $f(T,\mu_{\rm ch}-E)$ in $\tau_2$ for the purpose of variation with temperature and write,
\begin{align}
\Delta p(\tau_{\rm m}) &= [f(T,E-\mu_{\rm ch+})-f(T,E-\mu_{\rm ch-})]\times\nonumber\\ &\hspace{0.8cm}\left[1-\exp{ \left(-\frac{\tau_{\rm m}/\tau_{\rm a}}{\exp{(\Delta_2/k_{\rm B}T)}} \right)}\right].
\end{align}
The traps occupancy that actually dictates $\sigma_{\rm str}$ is when $V_{\rm g}$ is ramped back towards zero over time $\sim\tau_{\rm m}$ from the two extremes such that $\mu_{\rm ch}=\mu_{\rm th}$, see fig.\ref{fig:hyst-schem}. This will change $\Delta p$ by a factor $\sim\exp{ \left(-\frac{\tau_{\rm m}/\tau_{\rm a}}{\exp{(\Delta_2/k_{\rm B}T)}} \right)}$. This is similar to the super-paramagnets discussed in the Appendix. Different traps may have different energy barriers and with the fact that only traps in a narrow energy range near $\mu_{\rm ch}$ change their charge state, the distribution of $\Delta_2$ will dominate the temperature dependence of $V_{\rm th}$. Combining the weekly temperature dependent pre-factor $[f(T,E-\mu_{\rm ch+})-f(T,E-\mu_{\rm ch-})]$ with the slow trap's energy barrier distribution function $n_{\rm str}(\Delta_2)$, we conclude
\begin{align}
\Delta V_{\rm th}\propto \int n_{\rm str}(\Delta_2)&\left[1-\exp{\left(-\frac{\tau_{\rm m}/\tau_{\rm a}}{\exp{(\Delta_2/k_{\rm B}T)}} \right)}\right]\times \nonumber\\
&\exp{\left(-\frac{\tau_{\rm m}/\tau_{\rm a}}{\exp{(\Delta_2/k_{\rm B}T)}}\right)}d\Delta_2.
\label{eq:hyst-temp-evol}
\end{align}

In case of the same barrier value $\Delta_2$ for all traps we expect a peak in $\Delta V_{\rm th}$ with at a blocking temperature $T_{\rm B}=\Delta_2/[k_{\rm B}\ln(\tau_{\rm m}/\tau_{\rm a})]$. A distribution around a mean $\Delta_2$ will increase the width of this peak. Another unknown parameter here is $\tau_{\rm m}/\tau_{\rm a}$, \emph{i.e.} the ratio of measurement time, or $V_{\rm g}$ sweep time, to attempt rate.
\begin{figure}[h!]
	\centering
	\includegraphics[width=2.7 in]{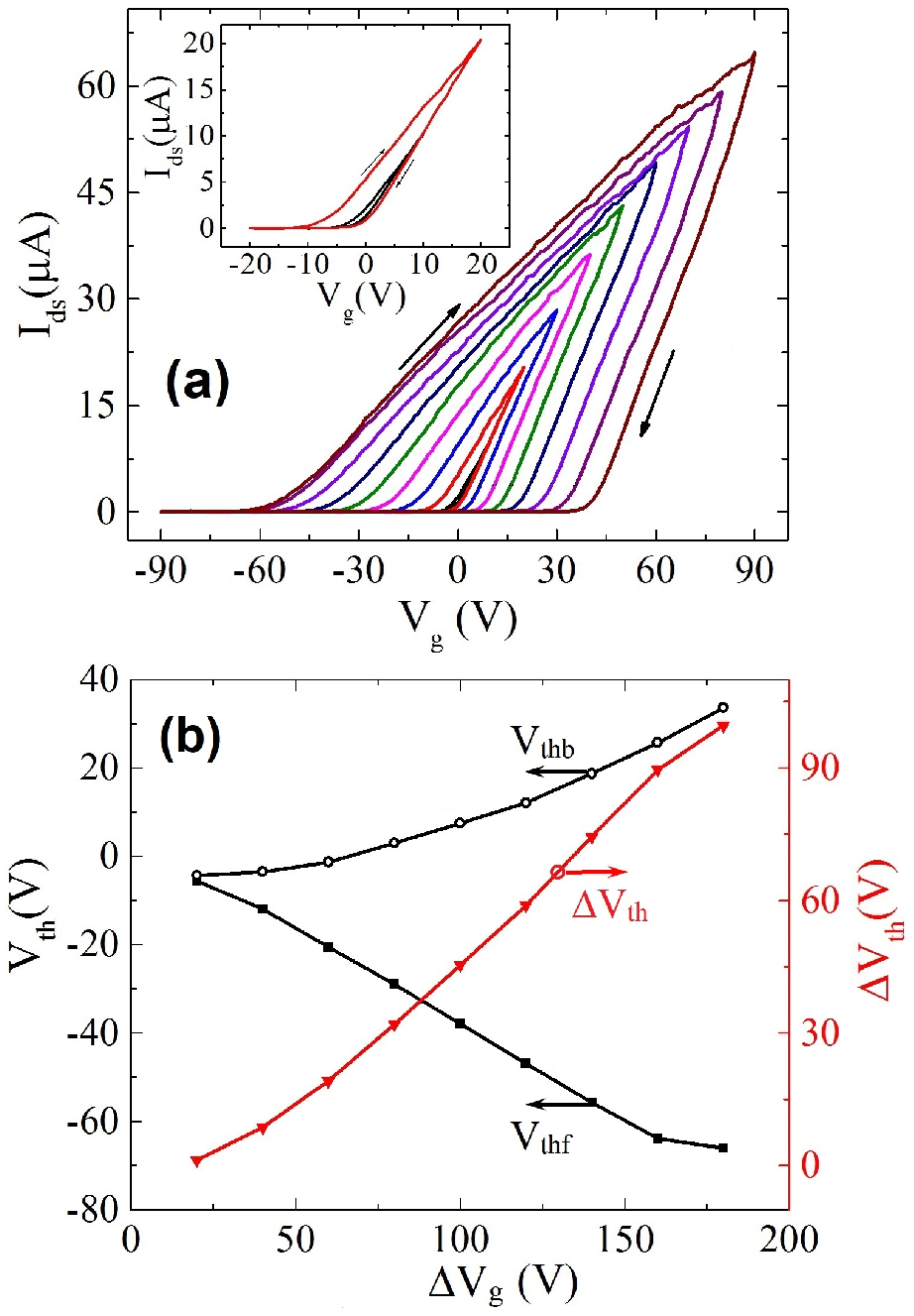}
	\caption{(a) $I_{\rm ds}$ Vs $V_{\rm g}$ at $V_{\rm ds}=1$ V for different sweep ranges of $V_{\rm g}$ from $\pm 10$ to $\pm 90$ V. All these curves are acquired at the same $V_{\rm g}$ sweep rate. The inset shows the zoomed-in portion for $\pm10$ and $\pm20$ V range $V_{\rm g}$ sweeps. (b) Variation of $V_{\rm thf}$, $V_{\rm thb}$ and $\Delta V_{\rm th}$ with sweep range $\Delta V_{\rm g}$ as extracted from  (a).}
	\label{fig:mos24}
\end{figure}
\section{IV: Experiments on hysteresis, blocking and gate-cooling}
In this section we discuss experimental measurements focusing on the slow traps in an FET device with a few-layer MoS$_2$ on SiO$_2$. This helps us understand the energy and barrier distribution associated with these traps. The observed temperature dependence of hysteresis, quantified by $\Delta V_{\rm th}$, is presented next together with the blocking model discussed earlier. Finally, the reversible handle on $V_{\rm th}$ through blocking of the traps in desired charge state by cooling under different gate voltages is discussed.
\subsection{IV-A: Hysteresis and time dependence at room temperature}
The transfer characteristics, shown in Fig. \ref{fig:hyst-schem}(a), of a few layer MoS$_2$ FET at room temperature exhibit a large hysteresis. The on-state high conductance at $V_g=+80$ V due to n-doping can be attributed to the electron rich sulfur vacancies and other n-type impurities present in natural MoS$_2$ crystals \cite{native defects}. This also leads to the pinning of $E_{\rm c}$ of MoS$_2$ close to the Fermi energy of contact-metal (gold) and thus negligible electron Schottky barrier at the MoS$_2$-metal contacts \cite{channel length,contact}. The blue line in Fig. \ref{fig:hyst-schem}(a) marks the subthreshold region for backward $V_{\rm g}$ sweep. The sub-threshold swing (SS) from this line works out as 3 V/dec as opposed to 0.06 V/dec, \emph{i.e.} the value expected for no traps, see Eq. \ref{eq:sub-th-swing}. This measured SS gives $\gamma_{\rm ftr}\approx50$ and $g_{\rm ftr}=3.7\times10^{12}$ eV$^{-1}$cm$^{-2}$.

Fig. \ref{fig:mos24}(a) shows the measured $I_{\rm ds}-V_{\rm g}$ curves for different $V_{\rm g}$ sweep ranges varying from $\pm10$ ($\Delta V_{\rm g}=20$) V, to $\pm90$ V ($\Delta V_{\rm g}=180$ V). There is negligible hysteresis for $\pm10$ V sweep range as $V_{\rm th}$ values for both the sweep directions are well within this sweep-range and nearly equal. With increasing sweep range, $V_{\rm thf}$ reduces and $V_{\rm thb}$ increases leading to a monotonic rise in $\Delta V_{\rm th}$, see Fig. \ref{fig:mos24}(b). Thus the slow traps, responsible for hysteresis, are nearly uniformly distributed over the $\mu_{\rm ch}$ range accessible up to the largest $V_{\rm g}$ sweep range. From $\Delta V_{\rm th}$ we can find the areal density of slow traps responsible for hysteresis for a given $\Delta V_{\rm g}$ by using $C_{\rm ox}\Delta V_{\rm th}/e$ with $C_{\rm ox}/e=7.6\times10^{10}$ ${\rm cm}^{-2}{\rm V}^{-1}$. Typical resulting values of areal density of slow traps $\sim10^{12}$ ${\rm cm}^{-2}$ are smaller than the usual three-dimensional (3D) semiconductors and similar to other 2D materials like graphene \cite{3d,graphene hysteresis}.

\begin{figure}[h!]
	\centering
	\includegraphics[width=2.7in]{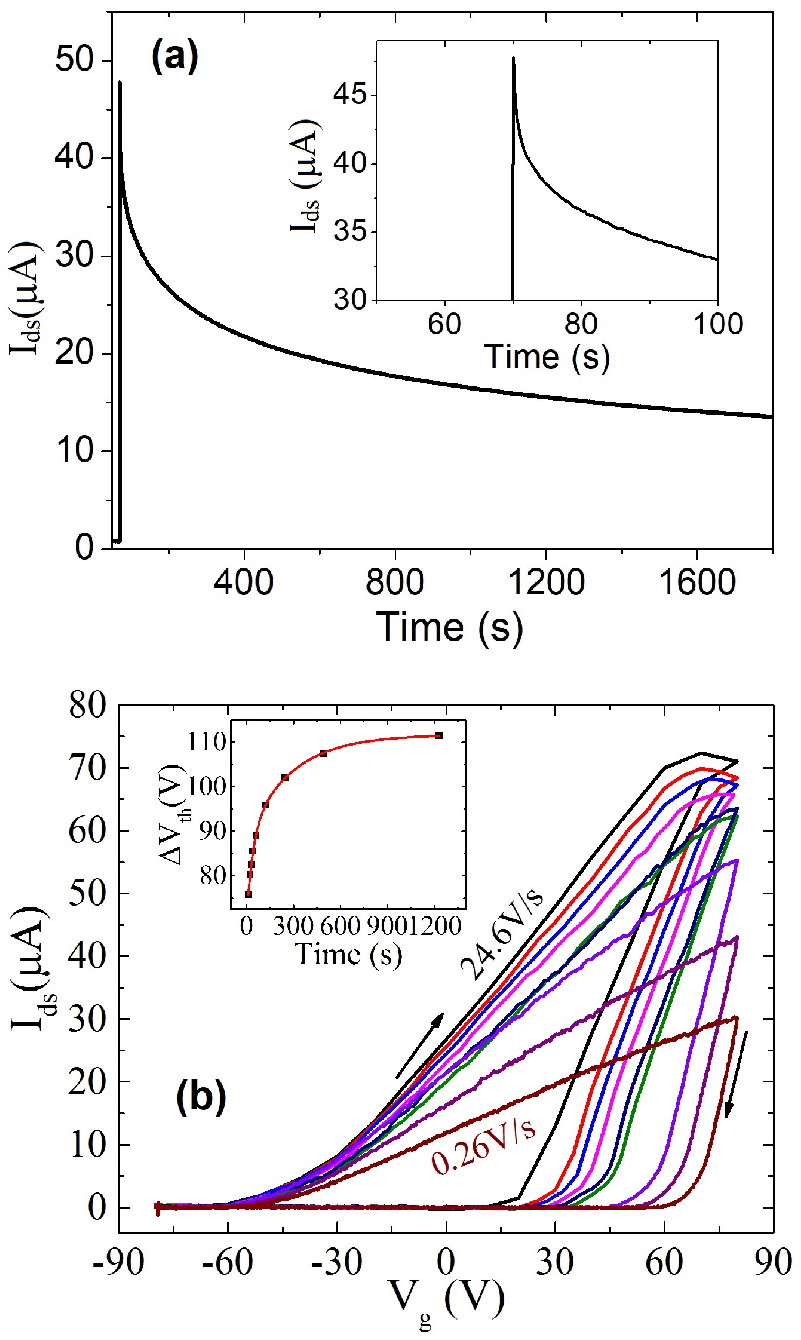}
	\caption{(a) The measured time dependence of $I_{ds}$ when $V_{\rm g}$ is changed abruptly from -80 to +80 V. The inset shows the zoomed-in initial part of the relaxation. (b) $I_{\rm ds}$ vs Gate voltage curves  for different sweep rates of back gate voltage at a fixed V$_{\rm ds}=1$ V. The solid squares in the inset show $\Delta V_{\rm th}$ as a function of overall sweep time with the red line showing a fit to a sum of two exponential relaxations.}
	\label{fig:mos23}
\end{figure}
A careful look at Fig. \ref{fig:mos24}(b) shows an asymmetry between $V_{\rm thf}$ and $V_{\rm thb}$ with the former changing more with $\Delta V_{\rm g}$ than later. This can be expected even for uniform distribution of traps as the magnitude of change in $\mu_{\rm ch}$ for positive $V_{\rm g}$ is less than that for negative $V_{\rm g}$. This is due to the rapid increase in channel's quantum capacitance when $\mu_{\rm ch}$ approaches $E_{\rm c}$. This will amount to activation of traps in narrower energy range for same the magnitude positive $V_{\rm g}$ than negative. A continuous rise in the rate at which $V_{\rm thb}$ changes with $\Delta V_{\rm g}$ and up to the highest $\Delta V_{\rm g}$ implies an increase in slow traps' DOS near $E_{\rm c}$. Also towards large $\Delta V_{\rm g}$ values $V_{\rm thf}$ seems to saturate indicating a reduction in slow trap's density of states when $\mu_{\rm ch}$ moves away from $E_{\rm c}$ and into the gap. From the monotonic rise in $\Delta V_{\rm th}$ with $\Delta V_{\rm g}$ we conclude that the slow traps are somewhat uniformly distributed. Although from the details of the $V_{\rm thf}$ and $V_{\rm thb}$ variation the traps seem to be concentrated over a limited energy range close to $E_{\rm c}$.

Figure \ref{fig:mos23}(a) shows a measured time dependent $I_{\rm ds}$ as a function of time when $V_{\rm g}$ is abruptly changed from $-80$ to +80 V. There is a fast initial relaxation followed by a slow stretched exponential tail indicating multiple time scales. This relaxation would be rather complex to fit to a microscopic model, as discussed earlier, in the absence of the knowledge about the distribution of trap energies and activation barriers. A fitting with multi-exponential or stretched exponential does work and it has indeed been used \cite{Bi-exponential} to conclude a distribution in barrier energies. However, due to the coupling between the dynamics of different trap's occupancy and $\mu_{\rm ch}$, even traps at single energy and with the same barrier can lead to non-exponential relaxation, with a long tail that can resemble a stretched-exponential, see Fig. \ref{fig:t-dep-ch}(b).

\begin{figure}[h!]
	\centering
	\includegraphics[width=3in]{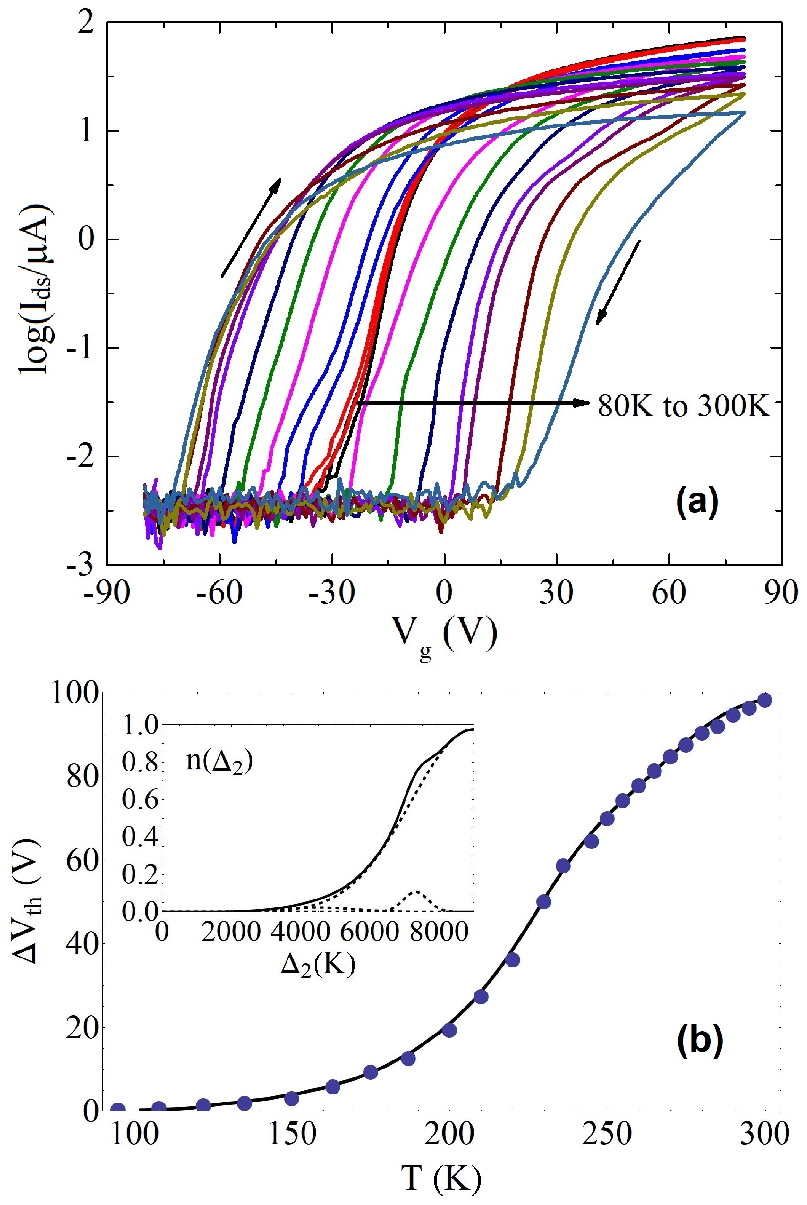}
	\caption{(a) Temperature dependence of $\log(I_{\rm ds})$ Vs $V_{\rm g}$ curves at $V_{\rm ds}=1$ V between 80 and 300 K. (b) The solid circles show $\Delta V_{\rm th}$ as a function of temperature. The solid line shows the calculated variation of $\Delta V_{\rm th}$ using Eq. \ref{eq:hyst-temp-evol} with the barrier distribution function $n(\Delta_2)$ depicted by the solid line in the inset. This $n(\Delta_2)$ is the sum of three Gaussian distributions shown by discontinuous lines in the inset. The traps with $\Delta_2$ beyond 8000 K do not contribute to $\Delta V_{\rm th}$ variation over the studied temperature range.}
	\label{fig:mos21}
\end{figure}
As a consequence of this slow relaxation of traps, the hysteresis has a significant dependence on the $V_{\rm g}$ sweep rate for a fixed sweep range. Fig. \ref{fig:mos23}(b) shows the conductance hysteresis loops acquired at different sweep rates from 0.26 to 24.6 V/s. A high sweep rate also gives higher peak conductance as less number of traps acquire negative charge leading to more electrons in the channel. In fact, for some of the very fast sweep rates, a saturation or a downturn in channel conductivity is seen with $V_{\rm g}$ due to a delayed response of the traps which depletes electrons from the channel. As discussed earlier, the rate of filling of an empty trap state at a given energy will increase with $V_{\rm g}$ as $\mu_{\rm ch}$ rises with $V_{\rm g}$. Fig. \ref{fig:mos24}(b) inset shows the variation of $\Delta V_{\rm th}$ as a function of the $V_{\rm g}$ sweep time. It fits well to a double exponential function, $\Delta V_{\rm th}=\alpha -\beta e^{-r_1\Delta t}-\gamma e^{-r_2\Delta t}$ with $r_{1}^{-1}=35$ s, $r_{2}^{-1}=292.5$ s and $\alpha$, $\beta$ and $\gamma$ as constants.

\subsection{IV-B: Blocking transition of interface traps}
Fig.\ref{fig:mos21}(a) shows $I_{\rm ds}-V_{\rm g}$ curves at several temperatures between 300 and 80 K over $\pm80$ V $V_{\rm g}$ sweep range and 2.6 V/s sweep-rate. For these measurements, the device was first kept at room temperature at $V_{\rm g}=0$ for 2-3 hours in order to equilibrate the traps and then cooled and stabilized at each different temperature keeping $V_{\rm g}=0$. The hysteresis can be seen to reduce with cooling though the rate of reduction is not monotonic as seen in Fig.\ref{fig:mos21}(b). $\Delta V_{\rm th}$ reduces slowly near room temperature and the rate of reduction, \emph{i.e.} $d\Delta V_{\rm th}/dT$, peaks near 225 K and then the rate as well as $\Delta V_{\rm th}$ diminish as 80 K temperature is approached.

When compared to superparamagnets, as discussed in the Appendix with parameters relevant to traps, one expects to see a peak in $\Delta V_{\rm th}$. In case of actual superparamagnets, where the attempt rate as well as barrier height are much smaller, the temperature dependence of M-H curves shows an opposite behavior as hysteresis disappears with increasing temperature. In that case, due to small barrier height the barrier can be made to vanish at accessible magnetic fields. However, in case of traps in MoS$_2$ devices the barrier is large and the accessible $V_g$ range only permits a small $\mu_{ch}$ variation which is insufficient to make the barrier vanish. Thus, at low temperatures, the traps do not change their state and one does not see hysteresis. Further, the hysteresis does not vanish at high temperatures, in case of traps, as there is a distribution in barrier height $\Delta_2$ which may continue to very high values for some of the traps.

The continuous line in Fig. \ref{fig:mos21}(b) shows the temperature dependence of $\Delta V_{\rm th}$ found using Eq. \ref{eq:hyst-temp-evol} and a $\Delta_2$ distribution depicted in the inset. Here we have used a fixed $\tau_m/\tau_0=10^{13}$ though a change in this value by up to even two orders of magnitude only slightly affects the required $n(\Delta_2)$ for fitting the measured $\Delta V_{\rm th}(T)$. Traps with $\Delta_2$ higher than 8000 K do not contribute to the hysteresis at temperatures 300 K or below. One may see a decline in $\Delta V_{\rm th}$ at further higher temperatures, however, we find that the $I_{\rm ds}-V_{\rm g}$ curves do not exhibit so sharp transitions at $V_{\rm thf}$ and $V_{\rm thb}$. This could be from activation of larger number of traps and some of the slow traps may turn into fast ones at higher temperatures. Other extrinsic effects, such as traps' diffusion, may also come into play. Eventually, very high $V_{\rm g}$ needed to access $V_{\rm thf}$ and $V_{\rm thb}$, and particularly at high temperatures, also leads to the breakdown of the dielectric oxide and permanent device damage.

\subsection{IV-C: Gate cooling and reversible control of $V_{\rm th}$}
\begin{figure}[h!]
	\centering
	\includegraphics[width=2.9in]{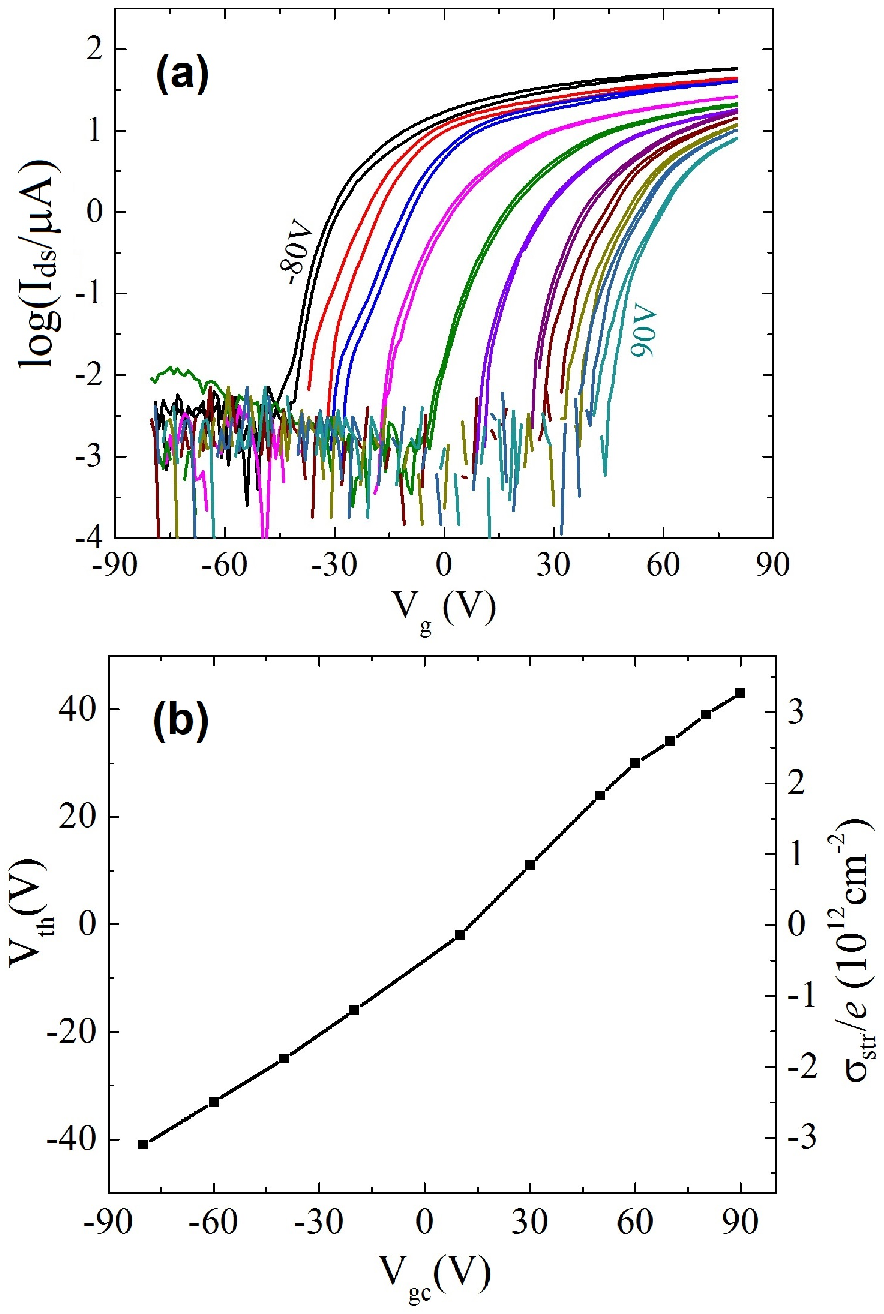}
	\caption{(a) Effect of cooling the device from 350 K to 80 K under different applied gate voltages $V_{\rm gc}$ from -80V to 90V. All the curves measured at 80 K and for $V_{\rm ds}=1$ V show negligible hysteresis. (b) Variation of $V_{\rm th}$ with $V_{\rm gc}$. The axis labels on the right in (b) show the corresponding blocked slow trap density.}
	\label{fig:mos22}
\end{figure}

Figure \ref{fig:mos22}(a) shows $I_{\rm ds}$ Vs $V_g$ measured at 80 K temperature after cooling the device from 350 K temperature to 80 K in presence of different gate voltages, labeled as $V_{\rm gc}$, between -80 and 90 V. The device was first warmed to 350K in vacuum and kept at desired $V_{\rm gc}$ for an hour before cooling it down to 80K. As expected there is negligible hysteresis at 80 K but more striking is the reversible change in $V_{\rm th}$ over a wide range from -40 to +40 V. At negative $V_{\rm gc}$ the traps get blocked in a positively charged state. This trap charge electron dopes the channel and thus a negative $V_{\rm g}$ is needed to deplete it. Similarly a positive $V_{\rm gc}$ leads to traps blocked with negative charge that depletes the electrons from the channel and thus a positive $V_{\rm g}$ is needed to make it conduct. In this way the traps act as a controllable virtual gate.

Figure \ref{fig:mos22}(b) shows the variation of $V_{\rm th}$ with $V_{\rm gc}$. The $V_{\rm th}$ value at 80 V can be converted into an appropriate charge density $\sigma_{\rm str}$ associated with the blocked slow-traps. The axis labels on right shows this $\sigma_{\rm str}/e=C_{\rm ox}V_{\rm th}$. Another fact from this figure is the nearly linear relation between $V_{\rm th}$ and $V_{\rm gc}$ with a slope close to 1/2. This indicates that about half of the charge induced by $V_{\rm gc}$ gets stored in the blocked slow-traps while the remaining half is taken up by the fast traps and channel carriers. This is striking as the change in $\mu_{\rm ch}$ (or $V_{\rm ch}$) with $V_{\rm g}$ near the conduction threshold is quite non-linear, see Fig.\ref{fig:with-trap-trans}.

\section{V: Discussion and conclusions}
When the thermally grown SiO$_2$ surface is stored in ambient air, the surface siloxanes terminated edge on the substrate react with water and gradually revert to Si-OH, after which the substrate becomes rehydrated which can act as electron trap center \cite{chemistry,chemistry1}. Further, a monolayer or submonolayer of hydrogen-bonded water stays on SiO$_2$ and cannot be removed by pumping in a vacuum even over long periods. This is also a possible source of interface traps. There can also be traps or dopants within MoS$_2$ channel that can arise from donor-like S mono-vacancies or more complex defects involving S vacancies. The slow traps having barriers of large heights, and presumably large widths, permit electron exchange only through thermal activation rather than through quantum tunneling. Other extrinsic effects where the change in interface charge happens through electrochemical reactions at the interface involving species of hydrogen and oxygen. This may bring the diffusion barriers for these species into the picture that can also influence the time scale of charge transfer.

Transport measurements are sensitive to the traps' distribution only over a narrow energy range near the conduction threshold. With a significant electron doping in MoS$_2$ on SiO$_2$ our results are consistent with a distribution of traps near $E_{\rm c}$. The inability to access hole-doped conduction even till $V_{\rm g}=-100$ V may indicate significant traps' density, both slow and fast, if one goes only by channel conduction. However, the contacts may also play a role and other investigations are required to conclude on this aspect. In fact, our attempts to access the hole doped regime by combining the gate-cooling at $V_{\rm gc}=90$ V and large negative $V_{\rm g}$ till -100 V at 80 K temperature also did not succeed.

In conclusion, a temperature dependent study of few-layer MoS$_2$ FET transfer characteristics shows hysteresis with a large difference $\Delta V_{\rm th}$ between the backward- and forward-sweep threshold gate-voltages. This is modeled using the hysteresis in interface trap charge density. The model also describes the complex coupled dynamics of channel carrier density and traps' charge density and thus even the traps with single energy and barrier can lead to non-exponential relaxations. The observed temperature dependence of $\Delta V_{\rm th}$ is attributed to the blocking of traps and fitted to a distribution of energy barriers for charge exchange between the traps and the channel. Finally, the blocking helps to get nearly non-hysteretic behavior at 80 K temperature with a voltage threshold programmable by gate-cooling voltage.

\section{Acknowledgements}
Authors acknowledge discussions on blocking transition with Ranjit Thapa and funding from SERB-DST of the Government of India.

\section{Appendix: Blocking in superparamagnets}
A superparamagnet consists of non-interacting nano-sized ferromagnetic crystals in single domain limit with their magnetic reversal described by the Stoner-Wohlfarth model \cite{stoner-wohlfarth}. Such crystals with uniaxial anisotropy exhibit two opposite magnetic moment $\pm m_{\rm s}$ states of equal energy, in the absence of external field, that are separated by an anisotropy energy-barrier, say $\Delta_0$. The external magnetic field $B$ along $+z-$direction makes spin-up state favorable to spin-down state by energy $2m_{\rm s}B$. Thus the barrier seen by spin-up state is increased to $\Delta_0+m_{\rm s}B$ and for the spin-down state it reduces to $\Delta_0-m_{\rm s}B$. We further assume the same time scale $\tau_0$ of magnetic dynamics in the two minima and the dominance of thermal activation over quantum tunneling. The rate of transition at temperature $T$ from spin-up to spin-down state will be given by $\tau_{\uparrow\downarrow}^{-1}=\tau_0^{-1}\exp[-(\Delta_0-m B)/k_{\rm B}T]$ and the reverse transition rate will be $\tau_{\downarrow\uparrow}^{-1}=\tau_0^{-1}\exp[-(\Delta_0+m B)/k_{\rm B}T]$. The average magnetic moment is given by $m=\langle m\rangle=(p_{\uparrow}-p_{\downarrow})m_{\rm s}$ with $p_{\uparrow}$ and $p_{\downarrow}$ as the probability of being in the respective spin-state. With $p_{\uparrow}+p_{\downarrow}=1$ we get $\langle m\rangle=(2p_{\uparrow}-1)m_{\rm s}$. The time dependence of $p_{\uparrow}$ is dictated by,
\begin{align*}
\frac{dp_{\uparrow}}{dt}=-\tau_{\uparrow\downarrow}^{-1}p_{\uparrow}+ \tau_{\downarrow\uparrow}^{-1}p_{\downarrow}= -(\tau_{\uparrow\downarrow}^{-1}+\tau_{\downarrow\uparrow}^{-1})p_{\uparrow} +\tau_{\downarrow\uparrow}^{-1}.
\end{align*}
This leads to the equation of motion for $m$ as,
\begin{align*}
\frac{dm}{dt}= -(\tau_{\uparrow\downarrow}^{-1}+\tau_{\downarrow\uparrow}^{-1})m+ (\tau_{\downarrow\uparrow}^{-1}-\tau_{\uparrow\downarrow}^{-1}).
\end{align*}
Substituting for $\tau_{\uparrow\downarrow}^{-1}$ and $\tau_{\downarrow\uparrow}^{-1}$, we get
\begin{align}
\frac{dm}{dt}= 2\tau_0^{-1}&\exp\left(-\frac{\Delta_0}{k_{\rm B}T}\right)\cosh\left(\frac{m_{\rm s} B}{k_{\rm B}T}\right)\times\nonumber\\&\hspace{2cm}\left[\tanh\left(\frac{m_{\rm s} B}{k_{\rm B}T}\right)-\langle m\rangle\right].
\label{eq:mag-dynamics}
\end{align}
Thus at equilibrium, \emph{i.e.} when $dm/dt=0$, we get $m_{\rm eq}=\tanh(m_{\rm s} B/k_{\rm B}T)$ expected for this two state system. However, the rate at which the equilibrium is attained is dictated by $2\tau_0^{-1}\exp(-\Delta_0/k_{\rm B}T)\cosh(m_{\rm s} B/k_{\rm B}T)$, which for $|m_{\rm s}B|>>k_{\rm B}T$ will become $\tau_0^{-1}\exp[-(\Delta_0\pm m_{\rm s}B)/k_{\rm B}T]$. Typical $\tau_0$ values for magnetic systems are of $\sim10^{-9}$ s order and thus within the measurement time scale $\tau_{\rm m}\sim1$ s, the equilibrium is attained either for $m_{\rm s}|B|\gtrsim\Delta_0$ or for $T\gtrsim T_{\rm B}=\Delta_0/[k_{\rm B}\ln(\tau_{\rm m}/\tau_0)]$. The former corresponds to the vanishing of the barrier between two states due to applied field and the latter defines the blocking temperature $T_{\rm B}$ with hysteresis for $T<T_{\rm B}$ and no hysteresis for $T>T_{\rm B}$.

\begin{figure}[h!]
	\centering
	\includegraphics[width=3.3 in]{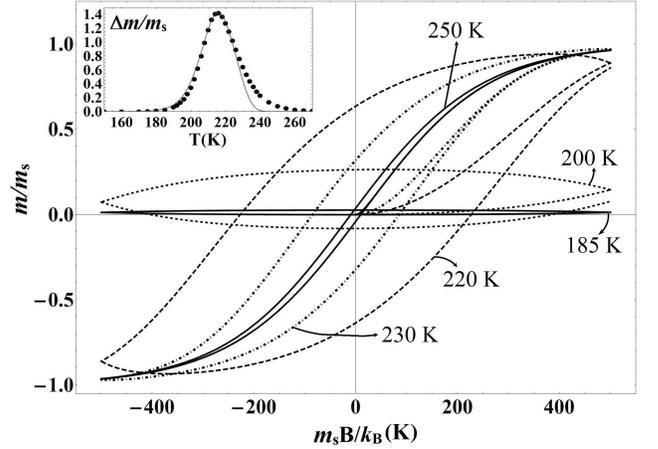}
	\caption{Calculated average magnetic moment ($m$) Vs applied field ($B$) at different temperatures. The field is swept in a cycle between $\pm 500 k_{\rm B}/m_{\rm s}$ at constant rate with total time $2\tau_{\rm m}$ (see text for details). The dots in the inset show the variation of the difference in two $m$ values at zero field, \emph{i.e.} $\Delta m$, with temperature illustrating how the hysteresis peaks near 217 K. The continuous line in inset is the plot following from Eq. \ref{eq:sup-par-analytical} with $a=0.76$ and $b=0.26$.}
	\label{fig:sup-par-model}
\end{figure}
In order to model the temperature dependence of hysteresis with parameters relevant to the charge traps, we consider a superparamagnetic-like system but with a large $\tau_0^{-1}$ and a large barrier $\Delta_0$ such that $\Delta_0>>m_{\rm s}B>>k_{\rm B}T$. Therefore, the barrier does not vanish at any typical applied $B$ and in fact the barrier always dominates the energetics. We can solve Eq. \ref{eq:mag-dynamics} numerically for a time-dependent $B$ changed at a constant rate and in a cycle. Starting from $t=0$, where $B=0$ and $\mu=0$, $B$ is ramped up to $+B_0$ over time $\tau_{\rm m}/2$. It is then ramped down and to $-B_0$ and then again to $+B_0$, all at the same rate. We assume $\Delta_0/k_{\rm B}=6500$ K, $m_{\rm s}B_0/k_{\rm B}=500$ K and $\tau_{\rm m}/\tau_0=10^{13}$. This leads to $m$ Vs $B$ as plotted in Fig. \ref{fig:sup-par-model}. At low temperatures (say, 185 K) we see that both the response of $m$ to the magnetic field and the hysteresis are negligible. As temperature rises, the response and hysteresis both increase but at high temperatures (say 250 K), the response is large but hysteresis vanishes. We use the difference $\Delta m$ in $m$ values at $B=0$ during reverse and forward field change of the same cycle as a measure of the hysteresis. As seen in the inset of Fig. \ref{fig:sup-par-model}, this $\Delta m$ exhibits a peak at the blocking temperature $T_{\rm B}= \Delta_0/[k_{\rm B}\ln(\tau_{\rm m}/\tau_0)]=217$ K.

The solid line in Fig. \ref{fig:sup-par-model} inset, given by
\begin{align}
\frac{\Delta m}{m_{\rm s}}=2&\left[1-\exp\left\{-\frac{\tau_{\rm m}}{2\tau_0}\exp\left(-\frac{\Delta_0 -am_{\rm s}B_0}{k_{\rm B}T}\right)\right\}\right]\nonumber\\
\times &\exp\left\{-\frac{\tau_{\rm m}}{2\tau_0}\exp\left(-\frac{\Delta_0 +bm_{\rm s}B_0}{k_{\rm B}T}\right)\right\},
\label{eq:sup-par-analytical}
\end{align}
follows from an analytical model behavior with $a$ and $b$ as constants between zero and one. The forward (higher to lower energy state) relaxation rate at field $B_0$ is $\tau_{\rm f}^{-1}=\tau_0^{-1}\exp[-(\Delta_0-m_{\rm s}B_0)/k_{\rm B}T]$ while the reverse rate is $\tau_{\rm r}^{-1}=\tau_0^{-1}\exp[-(\Delta_0+m_{\rm s}B_0)/k_{\rm B}T]$. The zero field rate will be $\tau_{\rm z}=\tau_0^{-1}\exp[-(\Delta_0)/k_{\rm B}T]$. As per Eq. \ref{eq:mag-dynamics}, when $B$ is changed abruptly to $B_0$ from zero, $m$ will change from zero to $m_1=\tanh(m_{\rm s}B_0/k_{\rm B}T)[1-\exp(-\tau_{\rm m}/2\tau_{\rm f})]$ in time $\tau_{\rm m}/2$. We can take $\tanh(m_{\rm s}B_0/k_{\rm B}T)=1$ for large $B_0$. Now if $B$ is abruptly made zero $m$ will become $m_2=\mu_1\exp(-\tau_{\rm m}/2\tau_{\rm z})$ after time $\tau_{\rm m}/2$. For continuous ramp the actual relaxation rate will be in between $\tau_{\rm f}^{-1}$ and $\tau_{\rm z}^{-1}$ for forward ramp and in between $\tau_{\rm z}^{-1}$ and $\tau_{\rm r}^{-1}$ for return ramp leading to $0<a,b<1$. A similar excursion in $B$ from zero to $-B_0$ will lead to $-m_2$ and thus $\Delta m=2m_2$.


\begin{thebibliography}
\bibitem{trap-band-info} In the schematics in Fig. \ref{fig:traps}, the band-system moves up or down depending on its potential and filling while $\mu_{\rm ch}$ remains fixed at zero similar to Fig.\ref{fig:Vg-band-shift}. The interface traps being in close vicinity of the channel will experience the same potential and thus the trap energy levels will move up or down with the bands. The channel bands will actually shift down (up) with increasing negative (positive) charge in the bands.
\end{thebibliography}
\end{document}